\documentclass[twocolumn]{jpsj2}

\def\runtitle{ }
\def\runauthor{Pinaki Majumdar  and Sanjeev Kumar}

\def\sf{\rm}

\title{
Structural Disorder Induced  Polaron Formation 
and Magnetic Scattering 
in the Disordered Holstein-Double Exchange Model}

\author{
Pinaki Majumdar  and Sanjeev Kumar}

\inst{
Harish-Chandra  Research Institute,\\
 Chhatnag Road, Jhusi, Allahabad 211 019, India
}

\abst{
In this paper we present results on the disordered Holstein-Double Exchange
model, explicitly in three dimension and `metallic' densities, obtained by
using a recently developed Monte Carlo approach. Following up on our earlier
paper, cond-mat 0406085, here we provide a detailed microscopic picture of 
the thermally driven metal-insulator transition (MIT) that arises close to 
the ferromagnet to paramagnet transition in this problem. This paper is 
focused mainly on the `diagnostics', clarifying the origin of the effective
disorder that drives the MIT in this system. To that effect, we provide 
results on the thermal evolution of the distributions of (i)~lattice 
distortions, (ii)~the net `structural disorder' and (iii)~the `hopping 
disorder' arising from spin randomness feeding back through the Hunds coupling. 
We suggest a phenomenology for the thermally driven MIT, viewing it as an 
`Anderson-Holstein' transition.
}

\kword{
colossal magnetoresistance manganites,
double exchange, lattice polarons, electron localisation
}

\begin{document}
\sloppy
\maketitle

\section{Introduction} 
The  Holstein-Double Exchange (H-DE) model has been the focus of
recent attention because it can help clarify 
the combined effect of strong electron-phonon
(EP) interaction and strong electron-spin coupling
that operates in the manganites, 
La$_{1-x}$Ca$_x$MnO$_3$, say \cite{mang-ref}.
The detailed  microscopic model for manganites is rather complicated.
Let us start by describing this model, and argue how successive reduction
to the disordered H-DE model retains some  of the essential physics.
 
The comprehensive tight binding model for perovskite manganites, 
neglecting oxygen orbitals, {\it etc}, is
\begin{eqnarray}
H &=& \sum_{\langle ij \rangle \sigma}^{\alpha \beta}
t_{ij}^{\alpha \beta} c^{\dagger}_{i \alpha \sigma}
c^{~}_{j \beta \sigma} + \sum_i \epsilon_i n_i
 - J_H\sum_i {\bf S}_i {\bf .} {\hat {\bf \sigma}_i }
\cr
&& - \lambda \sum_i {\bf Q}_i {\bf .} {\hat {\bf  \tau}_i}
+ H_\mathrm{stiff} + H_\mathrm{ph-dyn} + H_\mathrm{Hubb}
\end{eqnarray}
The  $t_{ij}^{\alpha \beta}$ are hopping between doubly degenerate
Mn $e_g$ levels at neighbouring sites,
$\epsilon_i$ refers to weak substitutional disorder,
and $J_H$ is the
strong Hunds coupling between the $e_g$ electrons
and the $S=3/2$ `core spin' ($t_{2g}$ electrons).
The `orbital moment', ${\bf \tau}_i$,
of the $e_g$ electrons is Jahn-Teller
coupled to the octahedral distortion parameter
${\bf Q}_i$, through the EP coupling $\lambda$.
$H_\mathrm{stiff}$  is  the stiffness of the lattice. In its
simplest form it is $\sim (K/2) \sum_i Q_i^2$, but
in reality  is of `cooperative' character, {\it i.e}, involves
phonon degrees of freedom
in  more than one octahedra.
The  phonons are quantum variables, with  intrinsic dynamics
arising from  $H_\mathrm{ph-dyn} \sim P^2/(2M_Q)$, where $P$ is the 
momentum and $M_Q$ the mass of the relevant oscillator.
Finally, the
$x=0$ state in the manganites is a Mott insulator, arising (partly) from
large on site `inter-orbital' Hubbard repulsion, $H_\mathrm{Hubb} \sim
U'\sum_{i, \alpha  \neq \beta} n_{i\alpha} n_{i \beta}$.

The electron-phonon, electron-spin and Hubbard interactions are all
large $\gg t$, where $t$ is the typical hopping scale in the
problem, and therefore beyond the range of perturbation theory.
The detailed model, unfortunately,
is far too complex for the present methods
of many body theory, if it  were to be handled in a realistic
three dimensional situation.  However, it is possible to
simplify the model somewhat, recognising that
(i)~at large doping of the Mott insulator, {\it e.g},
$x \sim 0.3-0.4$,
the Hubbard interaction probably does not
have a qualitative effect,  (ii)~the phonons are in the
adiabatic regime, with typical frequency $\omega_\mathrm{ph} \ll t$ and,
as a first approximation, we can explore  the {\it adiabatic limit}
$\omega_\mathrm{ph} =0$, and (iii)~the $S=3/2$ may be approximated
as a `classical' spin.
 
This still leaves us with the problem of (disordered) electrons
strongly coupled to (classical) phonon and spin degrees of
freedom. The qualitative effects in such a system include
(i)~magnetic order, typically  ferromagnetism, arising from
Hunds coupling and electron delocalisation,  (ii)~the
possibility of polaron formation if the EP coupling is
sufficiently large compared to the kinetic scale, and (iii)~the
phase competetion between different kinds of long range order.
These complexities, observed experimentally in the manganites,
 can be recovered already at the level of the Jahn-Teller-Double
Exchange (JT-DE) model, neglecting $H_\mathrm{ph-dyn}$ and $H_\mathrm{Hubb}$.

The JT-DE model was explored
by Millis {\it et al.}, \cite{millis-muller} within dynamical
mean field theory (DMFT), and later by Dagotto and coworkers
\cite{dag-mc} using real space Monte Carlo (MC) tecniques.
The DMFT study could demonstrate a thermally driven metal-insulator
transition (MIT),
as observed in La$_{0.7}$Ca$_{0.3}$MnO$_3$,
although only at $x=0.5$.  The MC studies,
on the other hand,
have gone a long way in clarifying the various phases in
the JT-DE model, but the severe finite size handicap
has prevented estimate of transport properties.
 
If we tentatively accept that the essential  physics in the manganites
arises from the general interplay of EP coupling and double
exchange, then the JT-DE model can be further reduced to
a one band version: the
Holstein-Double Exchange  model, defined further on.
Indeed, it  was studied early on
\cite{roder}, simultaneously with the JT-DE work of Millis {\it et al.},
within a mean field approximation, to clarify the effect of
EP coupling on the ferromagnetic $T_\mathrm{c}$. That work did
not address transport properties.  More recently
the H-DE model has been studied via
many body CPA approximations \cite{green} and through
direct Monte Carlo simulation \cite{hde-mc}.
The MC study does find a thermally driven MIT
but only at low carrier density. 
In addition, {\it all  these studies
neglect
the effect of quenched 
disorder} which should have a dramatic impact
\cite{emin-buss} in a system with strong EP coupling.
 
The effect of disorder in `manganite models' have in fact
been discussed \cite{dag-dis,furu-dis}, but the focus has
been on its impact on phase competetion and bicriticality.
Disorder can lead to nanoscale cluster coexistence near a
first order phase boundary \cite{coex-expts} and much of
the disorder related theory has concentrated on modelling
this phenomena.
While this is a vital issue, there is a {\it completely independent
} effect of disorder that shows up in systems with strong EP
coupling.
Disorder can induce polaron formation, even below the polaronic
threshold for `clean' systems, and drastically modify the
residual resistivity, the optical spectral weight, and in fact
the entire transport response. Manganite data testifying to these
effects abound in the literature, but there has been
no theoretical effort to understand these phenomena.

\section{Experiments in the Manganites}
In the manganites,
A$_x$A'$_{1-x}$MnO$_3$ say, the physics
is controlled by varying $x$, or
the mean A-A' site cation radius $r_A$, or
the cation disorder $\sigma_A$,
arising from size mismatch.
Since the parameters $r_A$ and $\sigma_A$ themselves depend on
$x$, it is best to focus on data at a fixed doping level.
In this spirit, experimenters have
used chemical variation at
fixed $x$, varying  both $r_A$ (which controls the electronic
bandwidth) and $\sigma_A$ \cite{saitoh,akahoshi}.
Pressure experiments, on the other hand involve variations
in the bandwidth,
keeping the disorder constant \cite{hwang,postorino}.
Finally, the effect of disorder has been explored, keeping
$x$ and $r_A$  fixed, varying only $\sigma_A$ \cite{attfield,maignan}.
 
The main conclusion of these studies is that
(i)~large $r_A$ systems, La$_{1-x}$Sr$_x$MnO$_3$,  are
  `canonical DE magnets', with a saturated ferromagnetic  ground state,
`metallic' resistivity at all temperature, and modest magnetoresistance 
(MR) near $T_\mathrm{c}$,
(ii)~reducing $r_A$, as in  La$_{1-x}$Ca$_x$MnO$_3$,
maintains a  saturated ferromagnetic  ground state, but there is a
thermally driven MIT near $T_\mathrm{c}$, with associated colossal magnetoresistance,
and a pseudogap in density of states (DOS), 
(iii)~further reduction
in $r_A$ as in La$_{1-x-y}$Pr$_y$Ca$_x$MnO$_3$, leads to a
non saturated ferromagnetic ground state,
mixed phase tendency,
nanoscale clusters (depending on disorder), a field driven insulator-metal
transition  at zero temperature, and large MR over a wide temperature range.
All these occur with only a {\it few percent change in $r_A$}.
 
These data  define the problem for a `global' description of the effects
seen in the manganites.
With small variation in electronic bandwidth (or inversely the EP coupling),
 $\sim 10 \%$, a theory
should be able to reproduce the suppression in  $T_\mathrm{c}$,
the dramatic changes in  transport character, reduction of
low frequency optical spectral weight,
sharply enhanced residual resistivity, and emergence of a
pseudogap features in the DOS.
{\it Simultaneously}
it should reproduce similar effects if the quenched disorder were
increased, 
remaining at fixed EP coupling and bandwidth.

In our earlier paper \cite{dhde1}
we had solved the disordered Holstein-Double
Exchange ($d$-H-DE) model and qualitatively reproduced all the
trends above (except magnetic phase competetion, which requires
additional antiferromagnetic coupling).
 In this paper we focus on a generic point in
parameter space, where there is a thermally driven MIT, and
try to provide a detailed microscopic picture for the transition.

\section{Model and Method}
The $d$-H-DE model, with classical spins and phonons, is defined by:
\begin{eqnarray}
H & =& -t\sum_{\langle ij \rangle \sigma}
c^{\dagger}_{i \sigma} c^{~}_{j \sigma}
+  \sum_{i }(\epsilon_i - \mu) n_i  -J_H\sum_i {\bf S}_i.{\sigma}_i
\cr
&& ~~~~~~~~~- \lambda \sum_i n_i  x_i
+ {K \over 2} \sum_i x_i^2  
\end{eqnarray}
 
Here the $t$ are nearest neighbour hopping on a simple cubic lattice, 
$\epsilon_i$
is the quenched binary  disorder, with value $\pm \Delta$, and
 $J_H$ is the Hunds coupling.
$\lambda$ is the
EP interaction, coupling electron density to the local distortion $x_i$, 
and 
$K$ is the stiffness of the lattice.
The basic parameters in the problem are $\Delta/t$,   $\lambda/t$,
electron density $n$, and temperature $T$.
We work in the limit $J_H/t \rightarrow \infty$.
 We also set $K=1$, and   measure energy, frequency, $T$, {\it  etc}, in
units of $t$.
 
For $J_H/t \rightarrow \infty$, the Hunds coupling acts as a constraint,
orienting the electron spin at a site parallel to the core spin (and
projecting out the anti-parallel component) leading to the following
spinless fermion model:
\begin{eqnarray}
H  & \equiv &
-\sum_{\langle ij \rangle }
(t_{ij}\gamma^{\dagger}_i \gamma_j ~+~h.c~)
+  \sum_{i } \xi_i n_i  + {K \over 2}
\sum_i x_i^2 \cr
t_{ij} &= &~~ t (\cos{\theta_i \over 2}\cos{\theta_j \over 2}
+ i~\sin{\theta_i \over 2}\sin{\theta_j \over 2}e^{i(\phi_i - \phi_j)}) \cr
~\cr
\xi_i & =&  \epsilon_i - \mu - \lambda x_i
\end{eqnarray}
The $\gamma, \gamma^{\dagger}$ are spinless fermion operators, corresponding
to the `parallel' spin projection of the original electrons.
$\xi_i$ is the net `potential' seen by the electrons, while the hopping
`disorder' is controlled by $\theta_i$ and
$\phi_i$,   the polar and azimuthal angles, respectively,
 of ${\bf S}_i$.

The problem involves strong coupling and disorder, and needs to handle
thermal fluctuations.
We use a recently developed Monte Carlo technique \cite{tca-ref}
based on a ``travelling cluster approximation'' (TCA) to anneal
the classical variables.  
This allows us to use system sizes 
$N \sim 10^3$ and
address  
the MIT in the~model.
 
After equilibriating the phonon and spin degrees of freedom for
a specified set of parameters, disorder realisation,
and $T$, we compute the following: (i)~transport
properties,  adapting a
scheme \cite{epl} which now uses electronic 
eigenfunctions 
in the annealed
background, and Fermi factors, for $T \neq 0$,  
(ii)~DOS, (iii)~the thermally averaged lattice distortion, as
well as (iv)~the distribution of net `structural disorder' and
hopping disorder, and (v)~spatially resolved information
on density distribution, $n_{\bf r}$, and  the 
magnetic
correlation, 
$f_{mag}({\bf r}) = \sum_{{\bf r}'}
\langle {\bf S}_{\bf r}.{\bf S}_{{\bf r}'}
\rangle $, 
where ${\bf r}$, ${\bf r}'$ are nearest neighbours.

\section{Results}
In our earlier, main paper \cite{dhde1}, we have provided the $n-T$ phase
diagram in the clean limit for various $\lambda$, while the transport,
spectral and optical properties were shown for $n=0.3$ varying the EP
coupling and disorder.
We demonstrated a thermally driven MIT over a wide parameter regime
in $\lambda-\Delta$ but did not have room to provide a detailed
microscopic picture of the `thermal disorder'
that drives the MIT. This paper focuses on a single, generic, point in
parameter space $n=0.3$, $\lambda=2.0$, $\Delta=0.6$
and tracks the $T$ dependence of the distribution of effective disorder
as well as the spatially inhomogeneous underlying state.
% -------------------------------------------------------------

%\vspace{1.6cm}

\begin{figure}
\begin{center}
\includegraphics[height=3.90cm,width=6cm]{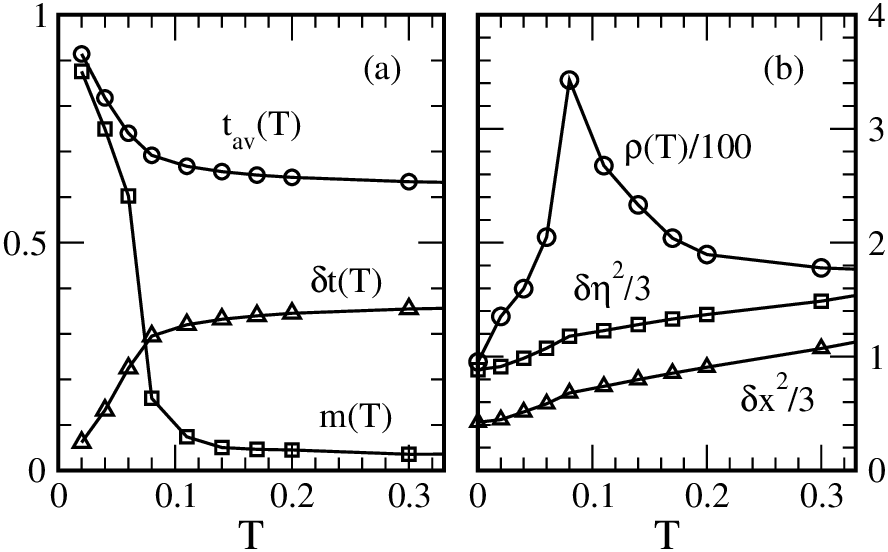}
\caption{Panel $(a).$~shows the variation in the magnetisation, $m(T)$,
the mean magnitude of the hopping amplitude, $t_{av}(T)$, and the
r.m.s fluctuation in the hopping, $\delta t(T)$. Panel $(b).$~shows
the
scaled resistivity, $\rho(T)$, the scaled variance of the lattice
distortion, $\delta x^2$, and the variance of the effective disorder
$\delta \eta^2$.}
\vspace{1.6cm}
%\end{center}
%\end{figure}
% -------------------------------------------------------------
%\begin{figure}
%\begin{center}
\includegraphics[height=4.0cm,width=7cm]{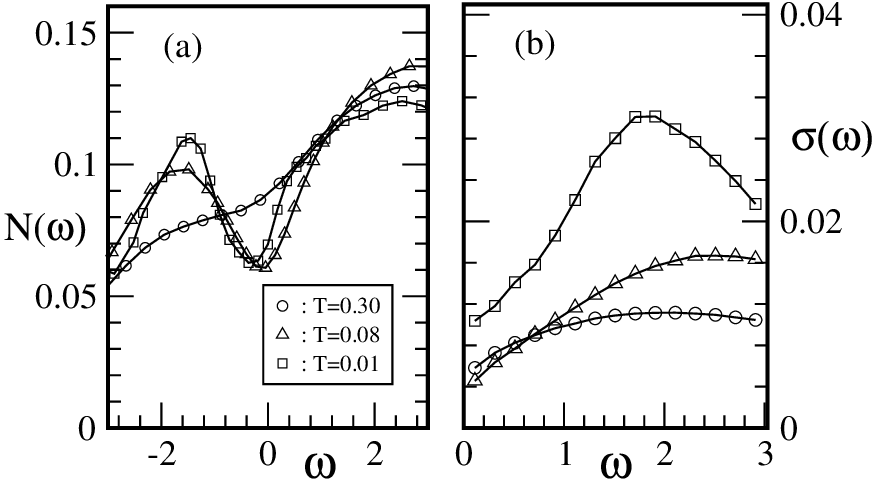}
\caption{(a)~The low energy density of states, and (a)~
the low frequency
optical conductivity, with varying temperature.}
\end{center}
\end{figure}
% -------------------------------------------------------------
%\newpage
% -------------------------------------------------------------
\begin{figure}
\begin{center}
\includegraphics[height=4.5cm,width=4.5cm]{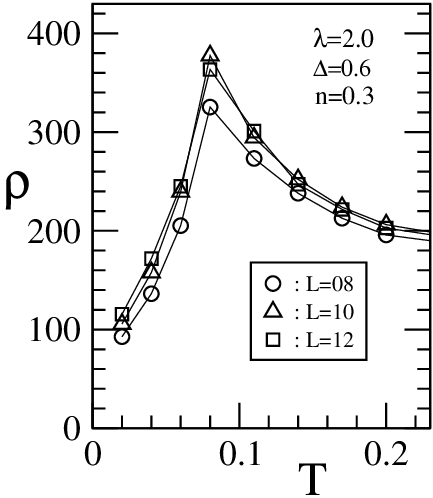}
\caption{The temperature dependence of the
`d.c' resistivity
at three system sizes, $L=8,~10,~12$, with $N=L^3$.
}
\vspace{1.5cm}
% -------------------------------------------------------------
\includegraphics[height=8.0cm,width=4.5cm]{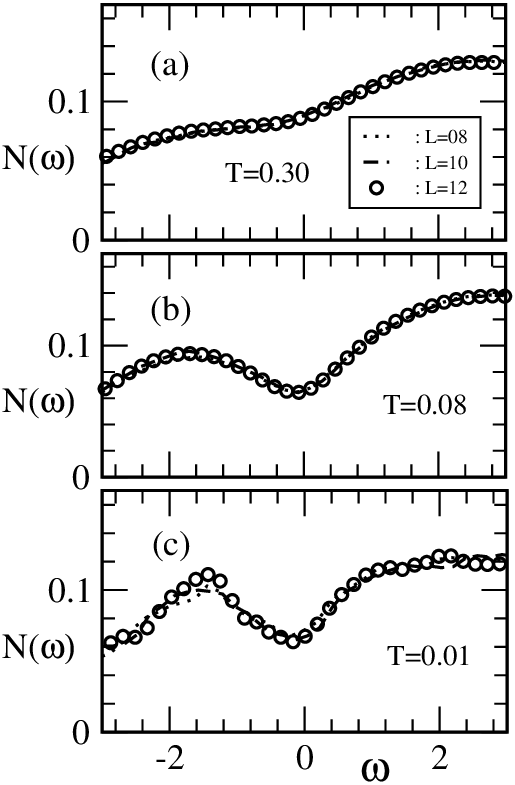}
\caption{Size dependence in the low frequency density of states,
showing the stability of the pseudogap feature. 
System sizes, $N=L^3$ with $L=8,~10,~12$, as in Fig.3.
}
\end{center}
\end{figure}
%\vspace{.3cm}

% -------------------------------------------------------------
If the thermally driven MIT in the manganites is a
``localisation transition'', rather than simple ``band splitting'', 
a real space understanding of the phenomena should be 
of crucial importance.

Figure~1(b) shows the resistivity at our chosen parameter point.
Figure~1(a) shows the variation
in magnetisation, $m(T)$, the spin disorder induced suppression
of the average,   $t_{av} = \langle \vert t_{ij}\vert \rangle$,
and the r.m.s fluctuation 
$\delta t(T)$ in the hopping distribution.
 
The evolution of the mean square 
lattice distortion, as well as the
effective structural disorder arising from $\epsilon_i - \lambda x_i$
are shown in panel (b). The full distribution of these disorder is
shown in Fig.~5.
The kink like feature in $\rho(T)$ is a consequence of a discrete set
of points sampled in $T$. Denser sampling reveals a more continuous
change in character.

Apart from the MIT, whose detailed description is our primary
focus in this paper, Fig.~2(a) shows the thermal evolution of
the low energy DOS, while Fig.~2(b) shows the low frequency
optical conductivity. The two features to note are (i)~the
non monotonic $T$ dependence of the low energy DOS (the `dip'
deepens initially with increasing $T$ and then fills up) and
(ii)~the strongly non Drude nature of the optical response
{\it even at $T=0$}. The resistivity and DOS had been shown
in our earlier paper \cite{dhde1}.  In what follows
we want to focus on the (annealed) disorder that is responsible
for these physical effects.
The finite size effects in this problem are rather weak, as borne
out by the size dependence of the resistivity in Fig.~3, and the
density of states, Fig.~4.

%\vspace{.2cm}
\vspace{\baselineskip}

\noindent
{\it Effects for $T \rightarrow 0$:}
Figure~1(b) shows that there are lattice distortions of
fairly large magnitude even at $T=0$, in the fully polarised
ferromagnetic state. An understanding of this can be obtained
by studying the disordered spinless  
Holstein model \cite{sk-pm-holst} 
where we observe that close to a (collective) polaronic
instability, even weak disorder can 
induce large lattice distortions and localise a 
fraction of charge carriers. This phase differs from a
polaronic insulator in that there are still extended states
that survive close to the Fermi level. The DOS shows signature
of this partial polaronic trapping ({\it i.e} a pseudogap),
and the optical response has the non Drude character typical of
electrons in a strongly disordered background.

Figure~5 shows the actual distributions,  of lattice distortions, $x_i$,
the net potential $\eta_i = \epsilon_i - \lambda x_i$, and
the magnitude of hopping, $\vert t_{ij}\vert$. The
quenched disorder is binary with value $\pm 0.6$.
Starting with panel (c), the hopping distribution $P(\vert t \vert)$
tends to a $\delta$ function as $T \rightarrow 0$ since all spins
are aligned. In this regime the physics is controlled purely
by the effective structural disorder.
The $P(x)$ distribution is  bimodal, with significant weight
near  $x \sim  \lambda/K = 2$, which would arise for 
strongly~localised~particles. 
There is also a peak 
related to the original 
clean Fermi liquid,  surviving near the origin.
% -------------------------------------------------------------
%\vspace{0.6cm}
\begin{figure}
\begin{center}
\includegraphics[height=6.5cm,width=4.7cm]{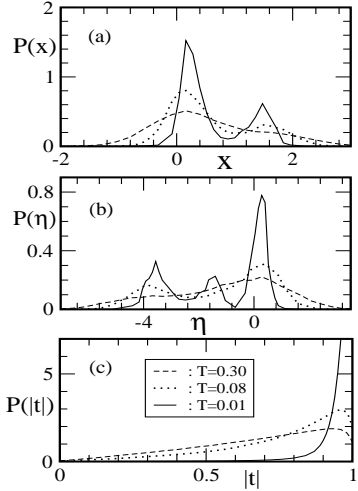}
\caption{
Panel (a): $P(x)$ the distribution of lattice distortions,
averaged thermally and over the system, (a)~the distribution 
of the effective potential, $\eta_i =
\epsilon_i - \lambda x_i$, seen
by the electrons, and (c)~the distribution in the magnitude 
of the hopping, essentially the nearest neighbour correlation
$\sqrt{ 1 + {\bf S}_i.{\bf S}_j}$.}
\end{center}
\end{figure}
% -------------------------------------------------------------
%\vspace{.1cm}
\begin{figure}
\begin{center}
\includegraphics[height=5cm,width=7cm]{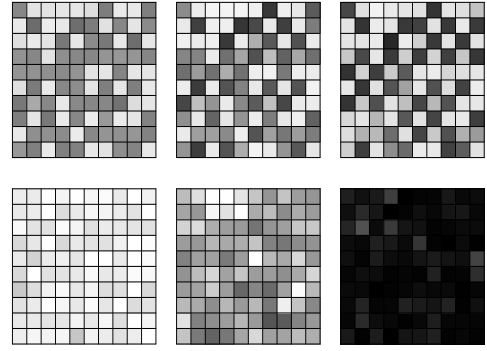}
\caption{The view on the top surface of a $10^3$ system.
The  row above  shows the thermally averaged density
profile, $n_{\bf r}$, at $ T = 0.01, 0.08, 0.30$ (right to
left). The row below shows the nearest neighbour averaged
magnetic correlation:
$f_{mag}({\bf r}) = \sum_{{\bf r}'}
\langle {\bf S}_{\bf r}.{\bf S}_{{\bf r}'}
\rangle $, 
for the same sequence of temperatures.}
\end{center}
\end{figure}
% -------------------------------------------------------------
 
We can crudely estimate the number of strongly localised carriers
in terms of $\int_{x_\mathrm{min}}^{\infty} dx P(x)$ where $x_\mathrm{min}=1.2$ 
is the
(approximate) upper limit of distribution at weak 
disorder,  $\Delta=0.2$, Fig.~7.
This area is roughly $0.2-0.25$
at $\Delta =0.6$.
Since the
electron density itself is $n=0.3$ this suggests that a large fraction
of carriers have collapsed into ``polaronic'' states. 
The density profile
at $T=0.01$, Fig.~6, top right, confirms this picture. That {\it not
all particles are localised} is borne out by the residual resistivity, 
$\rho(0)$. We have checked that large $L$ extrapolation of our results
(verified using  $N = 6^3-12^3$) still leads to finite $\rho(0)$.

The $P(\eta)$ distribution, at $T=0$, is more complex than  
$P(x)$, and is more relevant for understanding the 
`landscape' in which the electrons move. As a starting point we
can imagine three  kinds of sites: (i)~sites with $\epsilon_i 
=\pm \Delta$, with {\it weak distortions} and no `overdensity',
these sites would be continuation of the clean Fermi liquid, 
call them  $\eta_{FL}$,
(ii)~sites with $\epsilon_i = -\Delta$ and some large distortion
(call these $\eta^{-}$) and (iii)~sites with $\epsilon_i = +\Delta$
and moderately large distortions ($\eta^{+}$).

The $x_i$ for a `site localised' electron is $\sim \lambda/K =2$,
so the $\eta^{-}$ sites would have magnitude $\approx -\Delta
- \lambda^2/K = -4.6$. In fact the `polarons' are not quite site
localised, so $\eta^{-}$ would be somewhat less: consistent
with the left peak in Fig.~5(b). As for the $+\Delta$ sites,
these could have been avoided by the trapped particles if the
carrier density was low. 
However at $n=0.3$, some $+ \Delta$ sites
{\it also have distortions} and an associated overdensity, $\delta
n_{\bf r}$, but the 
magnitude is smaller than that for $-\Delta$ sites.
% -------------------------------------------------------------
\begin{figure}
\begin{center}
\vspace{0.0cm}
\includegraphics[height=5.5cm,width=6.0cm]{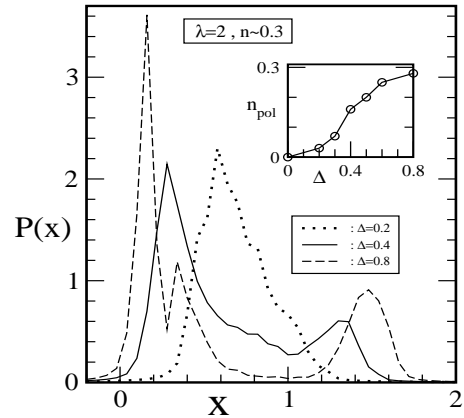}
\caption{The distribution $P(x, \Delta)$ at $T=0.01$ showing the
evolution of the polaronic peak with increasing disorder.
Inset: variation in the  density of localised carriers, $n_{pol}$, 
with increasing disorder, staying at $\lambda=2.0$ and fixed
electron density, $n=0.3$
We have used $n_{pol}(\Delta) = 
\int_{x_\mathrm{min}}^{\infty} dx P(x, \Delta)$ with  $x_\mathrm{min}=1.2$. 
}
\end{center}
\end{figure}
% -------------------------------------------------------------

This happens because 
locating the  localised particles always on $-\Delta$ sites would
sometimes require occupancy of adjacent sites. That would  inhibit the
(virtual) polaron hopping process, and lose kinetic energy. The
compromise is to lose some `potential energy' and put carriers
on non adjacent sites which may have unfavourable $\epsilon_i$.
This is the origin of the visible `checkerboard' pattern in 
Fig.~6, where particles generally prefer to remain on the diagonal
with respect to each other. 
At stronger disorder, pinning would dominate over kinetic energy 
considerations, and $+ \Delta$ sites would be depopulated.

The $P(\eta)$ 
distribution is  of course  not
$\delta$ functions at $\eta^{\pm}$ or $\eta_{FL}$ 
and only
has a broad three
peak character.

Based on the data discussed here, as well as the overall properties
presented earlier \cite{dhde1,sk-pm-holst} we suggest the 
following phenomenology for the $T=0$, spin polarised case:
(a)~The weak binary disorder tends to 
create a landscape where electronic
eigenfunctions, and the resulting density, $n_{\bf r}$, are
weakly inhomogeneous. The strong EP coupling amplifies this
inhomogeneity, by generating lattice distortions, leading to 
a state with a fraction of electronic states strongly localised.
Figure~7 shows $P(x)$ for several $\Delta$ at $T=0.01$, and also
the inferred density of polaronic states $n_{pol}$ at $n=0.3$.
(a)~Polaron formation 
transfers weight in $N(\omega)$ from around $\epsilon_F$
to lower energies, creating a pseudogap, and reduces the kinetic
energy (or effective carrier count), as visible in the low
frequency optical spectral weight \cite{dhde1}
$n_\mathrm{eff}({\bar \omega}, T)
= \int_0^{\bar \omega} \sigma(\omega, T) d\omega$,
by a factor of 10 (at ${\bar \omega} =1$)
with respect to the $\Delta=0$ probem.
(c)~There are, however, delocalised but strongly scattered 
states which still survive (at moderate $\Delta$): these give
rise to finite conductivity at $T=0$, and a `metallic' albeit
non Drude response in $\sigma(\omega)$.
$(d)$~In between the strongly localised `polaronic' states,
and the delocalised states near the Fermi level, there are
possible `Anderson localised' states as well, by which we
mean states with large localisation length, $\xi_{loc}$.
For `polaronic' states $\xi_{loc} \sim 1$.

A quantitative analysis in terms of these three kinds
of states is possible, by examining the electronic eigenfunctions
(or inverse participation ratio) in the annealed background. We
leave such discussion for the future.

%\vspace{.2cm}
\vspace{\baselineskip}

\noindent
{\it Effects at finite temperature:}
We have tried to argue how the interplay of disorder and
EP coupling leads to a strongly inhomogeneous state
at $T=0$. For ease of argument let us call this a 
 `two fluid' scenario, involving strongly localised
states below the Fermi level, and strongly scattered
extended states near $\epsilon_F$.
How does this picture evolve at finite $T$, when the 
magnetic degrees of freedom also come into play?

The resistivity itself increases quickly with
increasing temperature, Fig.~1(b), unlike what
is observed in the non magnetic case \cite{sk-pm-holst}
where the spin disorder does not feed back.
We argue that the localised states are now subject to
additional `hopping disorder', as evident in Fig.~1(a) and
Fig.~5(c), and remain ineffective in the conduction process.
In fact, they are localised marginally more due to the additional
randomness as the trend in the DOS reveals.

The extended states, which were earlier scattered by the potential
fluctuations, $\eta_i$, are now also scattered by the bond disorder
in $t_{ij}$ and experience reduced mobility. 
This situation is similar to the previously studied interplay of
structural disorder and paramagnetic scattering in the double exchange
model \cite{epl,lanl-loc}, where it was found that turning on 
spin disorder
indeed increases the resistivity of the structurally disordered system.
That problem, however, dealt with `quenched' structural and magnetic
disorder, while both of these are annealed variables in the present
problem.
Furthermore, 
as the magnetic disorder increases some of extended states
(or, more accurately, the fraction of extended states in a typical
equilibrium configuration) reduces. Depending on $\lambda$ and
$\Delta$, with increasing $T$ 
the combined disorder in $t_{ij}$ and $\epsilon_i 
-\lambda x_i$ (see eq.(3)) can drive the `mobility edge',
$\mu_\mathrm{c}(T)$,
above the Fermi level. This indeed happens at the parameter
point that we considered in this paper, while the 
full range of 
possibilities, including a simple `metal to metal' crossover,
is detailed in the previous paper \cite{dhde1}.

The                                                                                      high temperature
`insulating' phase that shows up for 
$T > T_{MIT} \sim T_\mathrm{c}$  
is different from a standard `Anderson insulator' for two reasons:
(i)~the localisation arises from a  
complex annealed disorder background, and not from an 
uncorrelated disorder distribution, hence the pseudogap
features, and (ii)~since the
thermal increase in disorder is itself driven by $T$,  
the mobility gap $\mu_\mathrm{c}(T) - \epsilon_F$ 
is always comparable to $T$, so there is no simple activated 
behaviour in the high temperature resistivity. This is roughly
consistent with what is observed in 
La$_{0.7}$Ca$_{0.3}$MnO$_3$, while the MIT is much stronger,
and the insulating phase more resistive,
in the PrCa systems.

\section{Discussion}
We have focused till now on our own results on the
MIT obtained in the adiabatic $d$-H-DE  model. 
It may be useful to place these results in relation to previous 
work exploring `bicriticality' in manganite models, and also comment
on cooperative lattice effects and quantum fluctuations in spins and
phonons.

(i)~The connection with  bicriticality: The issue of phase 
competetion,
the existence of first order phase boundaries, and the effect of
disorder in such a regime has been explored in simple
models \cite{dag-dis,furu-dis}. 
The issue was brought to focus by the early work of Dagotto
and coworkers \cite{dag-dis}, 
who suggested that the presence of disorder near a 
first order phase boundary between a ferromagnetic metal (FM) and an
antiferromagnetic (possibly charge ordered) insulator (AFI) could lead
to a pattern of coexisting FM and AFI clusters and percolative 
transport. The detailed transport properties in such a system,
probed within a microscopic theory, were clarified by us \cite{nano1}.
More recently a model involving Double Exchange, 
Holstein EP coupling, and disorder,
has been studied in a two dimensional system, at half-filling, and
discovered the ``metallisation'' of an intermediate coupling charge 
ordered (CO) phase by weak disorder. Our own results, far from a CO state,
suggest how a metal close to a polaronic instability is affected by the
presence of weak disorder.

Weak disorder has contrasting roles 
 in these different situations.
(a)~For a generic first order transition, between a FM and
an AFI, say \cite{dag-dis,nano1} disorder acts to convert ``macroscopic'' phase
separation to meso/nanoscale phase coexistence depending on the strength
of the random potential. 
 This harks back to the classic Imry-Ma scenario \cite{imry-ma}
and a fascinating variety of percolative effects can arise.
The key physical effect here is 
{\it disorder induced cluster formation near a 
first order transition} and does not involve phonons in any 
essential way.
In fact if phonons {\it were to be included} there are additional effects
of disorder (beyond Imry-Ma) which need to be considered. 
The second case \cite{furu-dis},
(a)~involves the effect observed in the  $n=0.5$ CO system. 
Here disorder generates a pinning potential, which acts as a random
field, and destroys the positional correlations of the CO state. Since
the intermediate coupling CO state depends on the periodicity of the
CO to generate insulating behaviour, destruction of CO promotes
metallisation \cite{furu-dis,sen-dag}, 
which we have observed in our 3d study
\cite{sk-pm-holst} as well. 
This is {\it disorder induced metallisation of an 
intermediate coupling CO state}.
If  the EP coupling were {\it large} then disorder could still 
destroy the CO phase but 
the charge ordered insulator would be converted to a 
charge disordered polaronic insulator. 
Our situation, (c)~corresponds
to a clean metal close to a polaronic instability being converted 
to a highly resistive state by the effect of disorder. Here {\it disorder
induces a polaronic instability}.  It should be clear that the
physical effect in (a) have no essential relation to those in 
(a) and (c).
While it is true that both
(a) and (c) arise in the same model, they occur on quite distinct
reference states, and the key weak disorder effects are physically 
very different, despite belonging to the same `global' $\lambda
-\Delta-n-T$  phase diagram.

(ii)~Order of the MIT: we think the `second order' character of our 
MIT is due to the effect of noncooperative phonons. If the lattice degrees
of freedom were directly coupled, as in the real material, then 
distortions at one site can have a cascading effect on the other sites,
generating an abrupt transition, as indeed observed in one MC study
\cite{hde-mc}.

(iii)~Quantum fluctuations in spins and phonons:
Our $T=0$ state has no quantum fluctuations (in spins or phonons).
While the quantum character of phonons affects the low $T$ resistivity
in clean metals, for the disordered strong coupling systems that we
consider (with large lattice distortions at $T=0$) quantum
fluctuations may not have a qualitative effect. It should be 
possible to make a perturbative estimate of the effects at small
phonon frequency $(\omega_\mathrm{ph}/t \ll 1)$, we have not done that 
yet. Similarly, the system is a fully polarised magnet at $T=0$.
Since the effective couplings in the problem are ferromagnetic we
do not anticipate a significant renormalisation in magnetic 
properties due to quantum  spin fluctuations.
Finally, by the 
time $T \sim T_\mathrm{c}$, where the MIT occurs,
 the thermal fluctuations in spins and phonons 
are far more important than quantum effects.

\section{Conclusion}
The role of disorder is crucial in the manganites, quite independent of
its effect at magnetic bicriticality. This is due to the inherently
strong EP coupling in the manganites, which places them close to
a polaronic instability. 
Disorder can partially 
trigger such an instability.
The interplay of disorder and EP coupling controls the resistivity
and the ferromagnetic $T_\mathrm{c}$. Even at fixed density and 
$r_A$ (or $\lambda$)
just increasing disorder can completely change the transport response.

While the present investigation has focused only on the thermally
driven metal-insulator transition, 
the combined effect of electron-phonon coupling, disorder, and
an antiferromagnetic coupling (to compete with double exchange)
can
generate responses that  interpolate between the 
`canonical DE magnets',
La$_{1-x}$Sr$_x$MnO$_3$, 
to thermally driven transitions,
La$_{1-x}$Ca$_x$MnO$_3$, 
all the way 
to bicriticality and
possible nanoscale coexistence,
as in La$_{1-x-y}$Pr$_y$Ca$_x$MnO$_3$.
A systematic use of our method should clarify
the origin, and provide a detailed microscopic 
description,
of the wide variety of transport regimes  
observed in the manganites.

%\vspace{.3cm}
\vspace{\baselineskip}

\noindent
We acknowledge use of the Beowulf cluster at HRI.

{}

\end{document}